\documentclass[12pt]{article}
\usepackage{epsf}
\usepackage{graphicx}
\usepackage{multicol}
\newcommand{\beq}{\begin{equation}}
\newcommand{\eeq}{\end{equation}}
\newcommand{\ba}{\begin{array}}
\newcommand{\ea}{\end{array}}
\newcommand{\beqa}{\begin{eqnarray}}
\newcommand{\eeqa}{\end{eqnarray}}

\newcommand{\cA}{{\cal A}}
\newcommand{\cO}{{\cal O}}
\newcommand{\cL}{{\cal L}}

\newcommand{\yuk}{{Y}}
\newcommand{\diagyuk}{{\lambda}}

\def\arnps#1#2#3{  {\it Annu. Rev. Nucl. Part. Sci. }{\bf #1}, #3 (#2)}
\def\npb#1#2#3{    {\it Nucl. Phys.}~B {\bf #1}, #3 (#2)}
\def\plb#1#2#3{    {\it Phys. Lett.}~B {\bf #1}, #3 (#2)}

\def\prd#1#2#3{    {\it Phys. Rev.}~D {\bf #1}, #3 (#2)}

\def\prl#1#2#3{    {\it Phys. Rev. Lett. }{\bf #1}, #3 (#2)}
\def\ptp#1#2#3{    {\it Prog. Theor. Phys. }{\bf #1}, #3 (#2)}

\def\ijmpa#1#2#3{  {\it Int. J. Mod. Phys. }~A {\bf  #1}, #3 (#2)}

\def\epjc#1#2#3{   {\it Eur. Phys. J.}~C {\bf #1}, #3 (#2)}
\def\jhep#1#2#3{   {\it JHEP  }{\bf #1}, #3 (#2)}

\def\ibid#1#2#3{{\bf #1}, #3 (#2)}

\begin{document}

\begin{center}
{\large\bf Kaon decays and the flavour problem}
\vskip 0.5 cm 
Gino Isidori \\
\vskip 0.3 cm 
{\em  INFN, Laboratori Nazionali di Frascati, Via E. Fermi 40, I-00044 Frascati, Italy }
\end{center}
 
\vskip 0.5 cm
\centerline{\large\bf Abstract}
\begin{quote}
After a brief introduction to the so-called {\em flavour problem},
we discuss the role of rare $K$ decays in probing the mechanism of 
quark-flavour mixing. Particular attention is devoted to the formulation of 
the {\em Minimal Flavour Violation} hypothesis, as a general and 
natural solution to the flavour problem, and 
to the fundamental role of $K \to \pi \nu {\bar\nu}$ decays
in testing this scenario.
\end{quote}

\vskip 0.7 cm 

\section{Introduction: the flavour problem}
Despite the Standard Model (SM) provides a successful description 
of particle interactions, it is natural to consider it only as the low-energy limit
of a more general theory, or as the renormalizable 
part of an effective field theory valid up to some 
still undetermined cut-off scale $\Lambda$. 
Since the SM is renormalizable, 
we have no direct indications about the value of $\Lambda$.  
However, theoretical arguments based on a natural solution of the 
hierarchy problem suggest that $\Lambda$ should not exceed a few TeV. 

One of the strategies to obtain additional clues about the
value of $\Lambda$ is to constrain (or find evidences) of the 
effective non-renormalizable interactions, suppressed by inverse powers 
of $\Lambda$, which encode the presence of new degrees of freedom
at high energies. These operators should naturally induce large effects 
in processes which are not mediated by tree-level SM amplitudes, 
such as $\Delta F=1$ and  $\Delta F=2$
flavour-changing neutral current (FCNC) transitions. 
Up to now there is no evidence of these effects and this 
implies severe bounds on the effective scale of dimension-six 
FCNC operators. For instance the good agreement between SM 
expectations and experimental determinations of $K^0$--${\bar K}^0$ 
mixing leads to bounds above $10^2$~TeV for the effective scale of 
$\Delta S=2$ operators, i.e. well above the few TeV 
range suggested by the Higgs sector. 

The apparent contradiction between these 
two determinations of  $\Lambda$ is a manifestation of what in 
many specific frameworks (supersymmetry, techincolour, etc.)
goes under the name of {\em flavour problem}:
if we insist with the theoretical prejudice that new physics has to 
emerge in the TeV region, we have to conclude that the new theory 
possesses a highly non-generic flavour structure. 
Interestingly enough, this structure has not been clearly identified yet,
mainly because the SM, {\it i.e.} the low-energy 
limit of the new theory, doesn't possess an exact flavour symmetry.
Then we should learn this structure from data, 
using the experimental information on FCNCs to constrain its form. 

\begin{figure}[t]
\begin{center}
$$
\includegraphics[width=10 cm, height=6 cm]{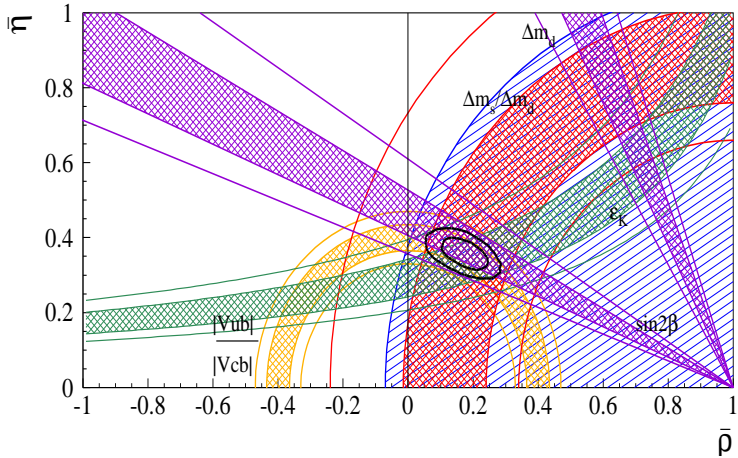}
$$
\end{center}
\caption{Allowed regions for the reduced Wolfenstein parameters \cite{Wolf}  
$\overline{\rho}$ and $\overline{\eta}$ (68\% and 95\% contours) 
compared with the uncertainty bands for $\left | V_{ub} \right |/\left | V_{cb} \right |$, 
$\epsilon_K$, $\Delta M_{B_d}$, the limit on $\Delta M_{B_s}/\Delta M_{B_d} $ and 
$a_{\rm CP}(B\to J/\Psi K_S)$ (from Ref.~\cite{Achille}). }
\label{fig:UT}
\end{figure}

Recently the flavour problem has been considerably exacerbated 
by the new precise data of $B$-factories, which show no 
sizable deviations from SM expectations also in  $B_d$--${\bar B}_d$ 
mixing and in a clean $\Delta B=1$ FCNC processes such as $B \to X_s \gamma$. 
One could therefore doubt about the need for new 
tests of the SM in the sector of (quark) flavour physics. 
However, there are at least two arguments why 
the present status cannot be considered conclusive 
and a deeper study of FCNCs is still very useful:
\begin{itemize}
\item
The information used at present to test the CKM mechanism \cite{CKM}
and, in particular, to constrain the unitary triangle, 
is obtained only from charged currents (i.e.~from tree-level amplitudes)
and $\Delta F=2$ loop-induced processes (see Fig.~\ref{fig:UT}). 
In principle, rare $K$ and $B$ decays mediated by $\Delta F=1$ FCNCs  
could also be used to extract indirect information 
on the unitary triangle, or to constrain new-physics effects. 
However, with the exception of the $B \to X_s \gamma$ rate, 
the quality of this information is very poor at present, 
either because of experimental difficulties or because 
of theoretical problems. 
Since new physics could affect in a very different 
way $\Delta F=2$ and $\Delta F=1$ loop-induced amplitudes
[e.g.~with $\cO(100\%)$ effects in the former and  $\cO(10\%)$
in the latter], it is mandatory to improve the quality 
of the $\Delta F=1$ information. 
\item
The most reasonable (but also most {\em pessimistic}) solution 
to the flavour problem is the so-called 
{\it Minimal Flavour Violation} (MFV) hypothesis. 
Within this framework, which will be discussed  
in detail in the next section, flavour-
and CP-violating interactions are linked to the
known structure of Yukawa couplings also beyond the SM. 
This implies that deviations from the SM 
in FCNC amplitudes rarely exceed the $\cO(10\%)$ 
level, or the level of irreducible theoretical 
errors in most of the presently available observables.
Moreover, theoretically clean quantities such as  
$a_{\rm CP}(B\to J/\Psi K_S)$ and $\Delta M_{B_d}/\Delta M_{B_s}$, 
which measure only ratios of FCNC amplitudes,  
turn out to be insensitive to new-physics effects. 
Within this framework the need for additional 
clean and precise information on  FCNC transitions
is therefore even more important. As we shall discuss 
in the following, the measurements of $\Gamma(K\to \pi \nu \bar \nu)$ 
would offer a unique opportunity in this respect
(see Ref.~\cite{YN} and references therein for a more extensive discussion).
\end{itemize}

\section{The Minimal Flavour Violation hypothesis}
The pure gauge sector of the SM is invariant under
a large symmetry group of flavour transformations: 
$G_F \equiv {\rm SU}(3)^3_q \otimes  {\rm SU}(3)^2_\ell\otimes U(1)^5$,
where 
${\rm SU}(3)^3_q  
= {\rm SU}(3)_{Q_L}\otimes {\rm SU}(3)_{U_R} \otimes {\rm SU}(3)_{D_R}$,
    ${\rm SU}(3)^2_\ell 
=  {\rm SU}(3)_{L_L} \otimes {\rm SU}(3)_{E_R}$
and three of the five $U(1)$ charges can be identified with 
baryon number, lepton number and hypercharge \cite{Georgi}. 
This large group and, particularly the ${\rm SU}(3)$ 
subgroups controlling flavour-changing transitions, is 
explicitly broken by the Yukawa interaction
\beq
\cL_Y  =   {\bar Q}_L \yuk_D D_R  H
+ {\bar Q}_L {\yuk_U} U_R  H_c
+ {\bar L}_L {\yuk_E} E_R  H {\rm ~+~h.c.}
\label{eq:LY}
\eeq
Since $G_F$ is broken already within the SM, 
it would not be consistent to impose it as an exact symmetry 
of the additional degrees of freedom
present in SM extensions: even if absent a the tree-level,
the breaking of $G_F$ would reappear at the quantum level 
because of the Yukawa interaction.  
The most restrictive hypothesis 
we can make to {\em protect} the breaking of $G_F$ 
in a consistent way, is to assume that 
$\yuk_D$, $\yuk_U$ and $\yuk_E$ are the only source of 
$G_F$-breaking also beyond the SM.

To implement and interpret this hypothesis in a natural way, 
we can assume that $G_F$ is indeed a good symmetry, promoting 
the $\yuk$ to be dynamical fields with 
non-trivial transformation properties under $G_F$:
\beq
\yuk_U \sim (3, \bar 3,1)_{{\rm SU}(3)^3_q}~,\qquad
\yuk_D \sim (3, 1, \bar 3)_{{\rm SU}(3)^3_q}~,\qquad
\yuk_E \sim (3, \bar 3)_{{\rm SU}(3)^2_\ell}~.
\eeq
If the breaking of $G_F$ occurs at very high energy scales 
 --- well above the TeV region where the new degrees of freedom 
necessary to stabilize the Higgs sector appear ---  at low-energies 
we would only be sensitive to the background values of 
the $\yuk$, i.e. to the ordinary SM Yukawa couplings. 
Employing the effective-theory language, 
we then define that an effective theory satisfies the criterion of
Minimal Flavour Violation if all higher-dimensional operators,
constructed from SM and $\yuk$ fields, are invariant under CP and (formally)
under the flavour group $G_F$ \cite{Georgi,noi}. 

According to this criterion one should in principle 
consider operators with arbitrary powers of the (adimensional) 
Yukawa fields. However, a strong simplification arises by the 
observation that all the eigenvalues of the Yukawa matrices 
are small, but for the top one, and that the off-diagonal 
elements of the CKM matrix ($V_{ij}$) are very suppressed. 
It is then easy to realize that, similarly to the pure SM case, 
the leading coupling ruling all FCNC transitions 
with external down-type quarks is \cite{noi}:
\beq\label{eq:FC}
(\lambda_{\rm FC})_{ij} = \left\{ \ba{ll} \left( \yuk_U \yuk_U^\dagger \right)_{ij}
\approx \lambda_t^2  V^*_{3i} V_{3j}~ &\qquad i \not= j~, \\
0 &\qquad i = j~. \ea \right.
\eeq
As a result, within this framework the bounds on 
the scale of dimension-six FCNC effective operators 
turn out to be much less severe than 
in the general case (see table~\ref{tab:tab}).
 
The idea that the CKM matrix rules the strength of FCNC 
transitions also beyond the SM has become a very popular 
concept in the recent literature and has been implemented 
and discussed in several works (see e.g. Refs.~\cite{MFV}). 
However, it is worth to stress that the CKM matrix 
represent only one part of the problem: a key role in
determining the structure of FCNCs  is also played  by quark masses, 
or by the Yukawa eigenvalues. In this respect the above MFV 
criterion provides the maximal protection of FCNCs (or the minimal 
violation of flavour symmetry), since the full structure of 
Yukawa matrices is preserved. We finally emphasize that, 
contrary to other approaches, the above MFV criterion 
is based on a renormalization-group-invariant symmetry argument 
and is completely independent of specific new-physics framework.

\renewcommand\arraystretch{1.2}
\begin{table}
$$
\begin{array}{lc|ccc}
\multicolumn{1}{c}{\hbox{Minimally flavour violating}} &\hbox{main}
&\multicolumn{2}{c}{\Lambda\hbox{ [TeV]}}&\\
\multicolumn{1}{c}{\hbox{dimension~six~operator}} &
\hbox{observables} &
-& +\\
\hline
 \frac{1}{2} (\bar Q_L  \lambda_{\rm FC} \gamma_{\mu} Q_L)^2 
\phantom{X^{X^X}}
&\epsilon_K,\quad \Delta m_{B_d}      &6.4 & 5.0 \\
    e H^\dagger \left( {\bar D}_R  \diagyuk_d \lambda_{\rm FC} \sigma_{\mu\nu}
Q_L \right) F_{\mu\nu}   &
B\to X_s \gamma     &5.2 & 6.9 \\
  (\bar Q_L  \lambda_{\rm FC}\gamma_{\mu}   Q_L)(\bar L_L \gamma_\mu L_L )  
&B\to (X) \ell\bar{\ell},\quad K\to \pi \nu\bar{\nu},(\pi) \ell \bar{\ell} \quad
  & 3.1 & 2.7  & *\\
  (\bar Q_L  \lambda_{\rm FC} \gamma_{\mu} Q_L)(H^\dagger i D_\mu H)\qquad  
&B\to(X) \ell\bar{\ell},\quad K\to \pi \nu\bar{\nu},(\pi) \ell \bar{\ell} \quad
 &  1.6 & 1.6 & *\\
\end{array}$$
\caption[X]{\label{tab:tab}  $99\%$ {\rm CL} bounds on the scale
of representative dimension-six operators in the MFV scenario \cite{noi}.
The constraints are obtained on the single operator, with coefficient $\pm1/\Lambda^2$
($+$ or $-$ denote constructive or destructive interference with the SM amplitude). 
The $*$ signals the cases where a significant increase in sensitivity 
can be achieved by future measurements of rare decays.
}
\end{table}
\renewcommand\arraystretch{1}

\section{$K\to\pi\nu\bar{\nu}$ decays}

\begin{figure}[t]
$$
\includegraphics[width=10cm]{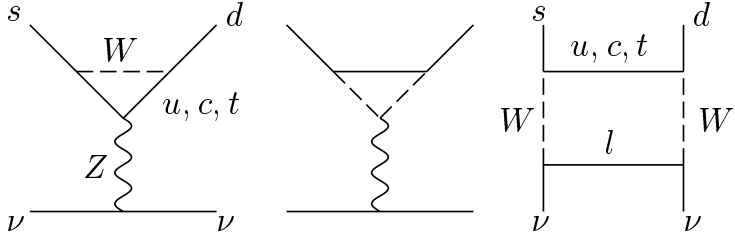}
$$
\caption{One-loop diagrams contributing to
 the $s \to d \nu \bar{\nu}$ transition.}
\label{fig:Kpnn}
\end{figure}

The $s \to d \nu \bar{\nu}$ transition is one 
of the rare examples of weak processes whose 
leading contribution starts at $\cO(G^2_F)$. At the one-loop 
level it receives contributions only from $Z$-penguin and 
$W$-box diagrams, as shown in Fig.~\ref{fig:Kpnn}, 
or from pure quantum electroweak effects.
Separating the contributions to the one-loop amplitude according to the 
intermediate up-type quark running inside the loop, we can write
\beqa 
&& \cA(s \to d \nu \bar{\nu}) = \sum_{q=u,c,t} V_{qs}^*V_{qd} \cA_q  
 \quad \sim \quad \left\{ \begin{array}{ll} \cO(\lambda^5
m_t^2)+i\cO(\lambda^5 m_t^2)\    & _{(q=t)}   \\
\cO(\lambda m_c^2 )\ + i\cO(\lambda^5 m_c^2)     & _{(q=c)} \\
\cO(\lambda \Lambda^2_{\rm QCD})     & _{(q=u)}
\end{array} \right. \quad
\label{uno}
\eeqa
where $V_{ij}$ denote the elements of the CKM matrix. 
The hierarchy of these elements 
would favour  up- and charm-quark contributions;
however, the {\em hard} GIM mechanism of the perturbative calculation
implies $\cA_q \sim m^2_q/M_W^2$, leading to a completely 
different scenario. As shown on the r.h.s.~of (\ref{uno}), 
where we have employed the standard CKM phase convention 
($\Im V_{us}=\Im V_{ud}=0$) and 
expanded the $V_{ij}$ in powers of the 
Cabibbo angle ($\lambda=0.22$), 
the top-quark contribution dominates both real and
imaginary parts.
This structure implies several interesting consequences for
$\cA(s \to d \nu \bar{\nu})$: it is dominated by short-distance
dynamics, therefore its QCD corrections are small and calculable in perturbation theory; 
it is very sensitive to $V_{td}$, which is one of the less constrained CKM matrix elements;
it is likely to have a large CP-violating phase; it is very suppressed within the SM and thus 
very sensitive to possible new sources of quark-flavour mixing.

Short-distance contributions to $\cA(s \to d \nu \bar{\nu})$,
both within the SM and within MFV models, can efficiently be described 
by means of a single effective dimension-6 operator:
$Q^{\nu}_{L}= \bar{s}_L\gamma^\mu d_L~\bar{\nu}_L \gamma_\mu \nu_L~.$
Within the SM both next-to-leading-order (NLO) QCD corrections \cite{BB,MU}
and $\cO(G_F^3 m_t^4)$ electroweak corrections \cite{BBmt} to the 
Wilson coefficient of $Q^{\nu}_{L}$ have been calculated.
The simple structure of $Q^{\nu}_{L}$ leads to two important  
properties of $K\to \pi \nu\bar \nu$ decays:
\begin{itemize}
\item{} The relation between partonic and hadronic amplitudes 
is exceptionally accurate, since hadronic matrix elements
of the $\bar{s} \gamma^\mu d$ current between a kaon and a pion
can be derived by isospin symmetry from the measured $K_{l3}$ rates. 
\item{} The lepton pair is produced in a state of definite CP 
and angular momentum, implying that the leading SM contribution 
to $K_L \to \pi^0  \nu \bar{\nu}$ is CP-violating.
\end{itemize}

\medskip 

The dominant theoretical error in estimating 
the $K^+\to\pi^+ \nu\bar{\nu}$ rate is due to the subleading, 
but non-negligible charm contribution. Perturbative NNLO corrections 
in the charm sector have been estimated 
to induce an error in the total rate of around  $10\%$ \cite{BB}, 
which can be translated into a $5\%$ error in the determination of 
$|V_{td}|$ from ${\cal B}(K^+\to\pi^+ \nu\bar{\nu})$. 
Non-perturbative effects introduced 
by the integration over charmed degrees of freedom
have been discussed in Ref.~\cite{Falk_Kp}: a precise 
estimate of these contributions is not possible, but 
they are expected to be within the error of NNLO terms.
Genuine long-distance effects associated 
to light-quark loops have been shown to be much smaller \cite{LW}.

The case of $K_L\to\pi^0 \nu\bar{\nu}$ is even cleaner from the
theoretical point of view \cite{Litt}.  
Because of the  CP structure, only the imaginary parts in (\ref{uno}) 
--where the charm contribution is absolutely negligible--
contribute to $\cA(K_2 \to\pi^0 \nu\bar{\nu})$. Thus 
the dominant direct-CP-violating component 
of $\cA(K_L \to\pi^0 \nu\bar{\nu})$ is completely 
saturated by the top contribution, 
where  QCD corrections 
are suppressed and rapidly convergent. 
Intermediate and long-distance effects in this process
are confined only to the indirect-CP-violating 
contribution \cite{BB3} and to the CP-conserving one \cite{CPC},
which are both extremely small.
Taking into account the isospin-breaking corrections to the hadronic
matrix element \cite{MP}, we can write an
expression for the $K_L\to\pi^0 \nu\bar{\nu}$ rate in terms of 
short-distance parameters, namely
\beq
{\cal B}(K_L\to\pi^0 \nu\bar{\nu})_{\rm SM}~=~4.16 \times 10^{-10} \times \left[
\frac{\overline{m}_t(m_t) }{ 167~{\rm GeV}} \right]^{2.30} \left[ 
\frac{\Im(V^*_{ts} V_{td})}{ \lambda^5 } \right]^2~, 
\eeq
which has a theoretical error below $3\%$.

\begin{figure}[t]
$$
\hskip -0.8 cm
\includegraphics[width=10.0cm,height=6.5cm]{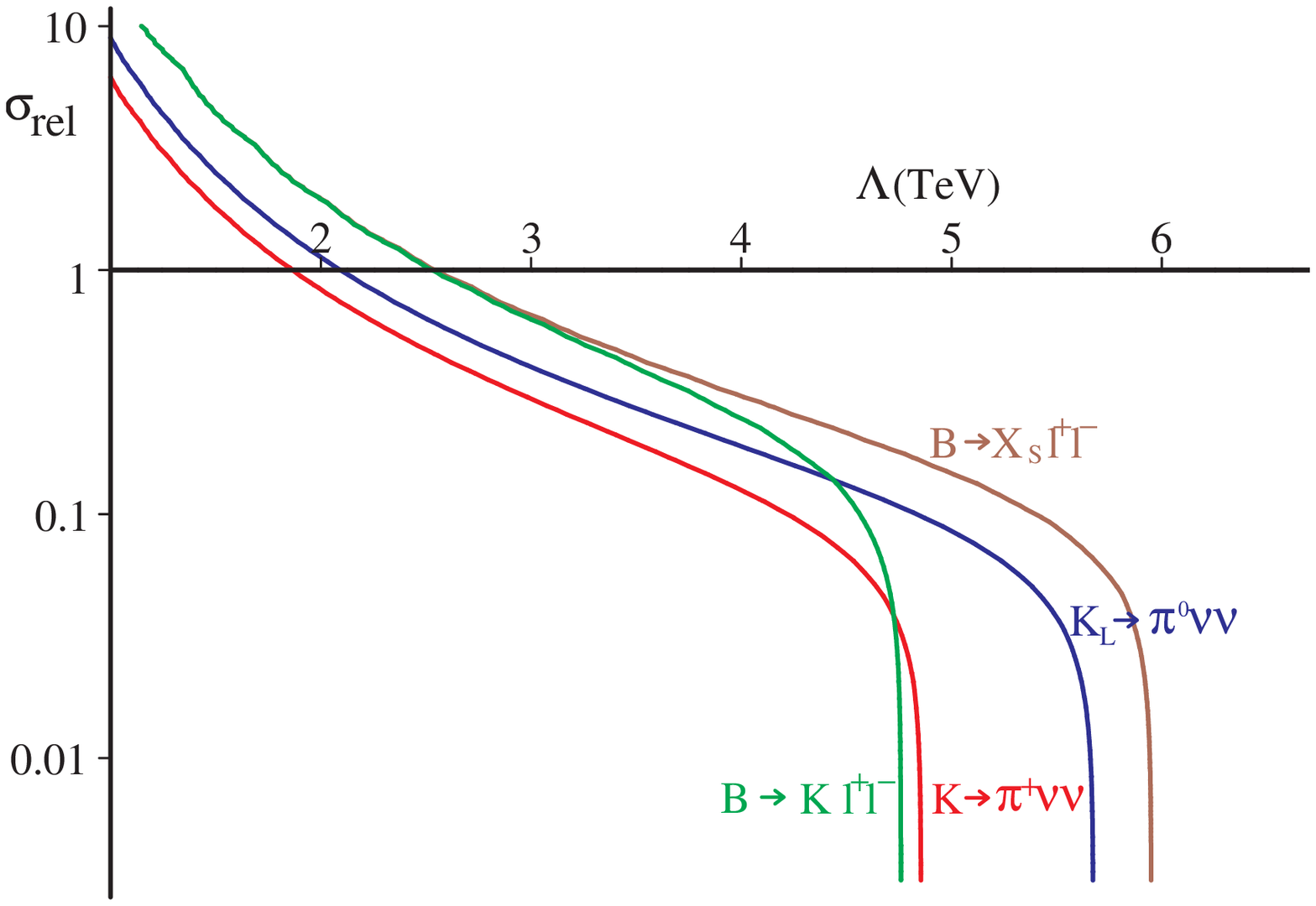}
\hskip -1.6 cm
\includegraphics[width=10.0cm,height=6.5cm]{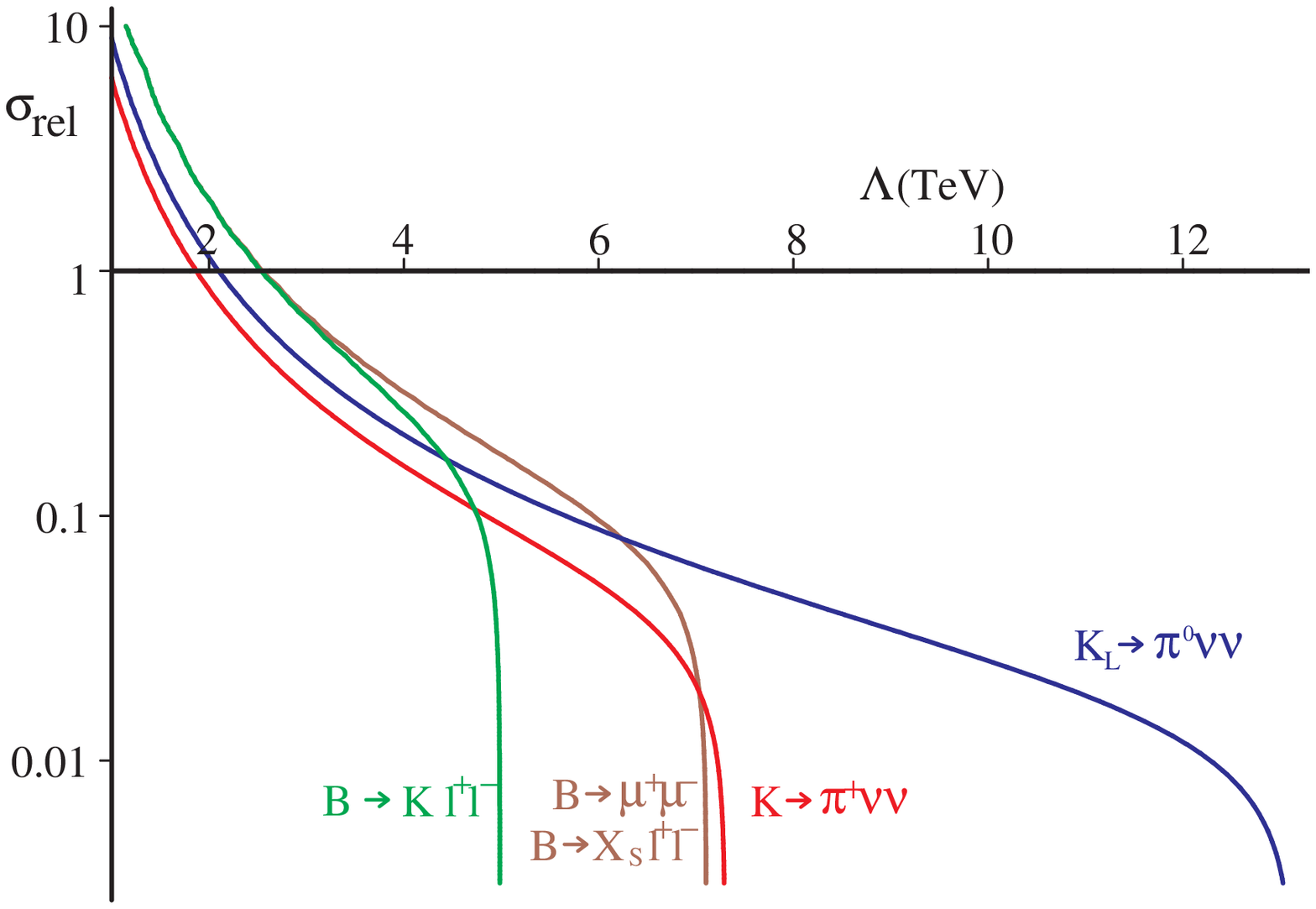}
\hskip -1.2 cm
$$
\vskip -1 cm
\caption[X]{\label{fig:conf} Comparison of the effectiveness 
of different rare modes  in setting future bounds 
on the scale of the representative operator 
$(\bar Q_L  \lambda_{\rm FC}\gamma_{\mu}   Q_L)(\bar L_L \gamma_\mu L_L)$
within MFV models \cite{noi}. The vertical axis indicates the relative precision 
of an hypothetic measurement of the rate, with central value equal to
the SM expectation. The curves in the two panels are obtained
assuming an uncertainty of 10\% (left) or 1\% (right) on the
corresponding overall CKM factor. }
\end{figure}

The high accuracy of the theoretical predictions of ${\cal B}(K^+ \to\pi^+
\nu\bar{\nu})$ and ${\cal B}(K_L \to\pi^0 \nu\bar{\nu})$ in terms of modulus
and phase of $\lambda_t=V^*_{ts} V_{td}$ clearly offers
the possibility of very interesting tests of flavour dynamics.
Within the SM, a measurement of both channels would provide, 
two independent pieces of information on the unitary triangle, 
or a complete determination of $\bar\rho$ and $\bar\eta$ from $\Delta S=1$
transitions. As illustrated in  Fig.~\ref{fig:Kpnn},
the high sensitivity of these two modes to short-distance dynamics
makes them extremely efficient probes of MFV scenarios.

At present the SM predictions of the two  $K\to \pi\nu\bar\nu$ rates
are not extremely precise owing to the limited knowledge of $\lambda_t$. 
Taking into account all the indirect constraints in a global Gaussian fit, 
the allowed range reads \cite{rising,kettel}
\beqa
{\cal B}(K^+ \to\pi^+ \nu\bar{\nu})^{ }_{\rm SM} &=& (0.72 \pm 0.21) 
\times 10^{-10} \qquad  \label{BRK+nnt}
\\
{\cal B}(K_L \to\pi^0 \nu\bar{\nu})^{ }_{\rm SM} &=& (0.28 \pm 0.10) 
\times 10^{-10} \qquad  \label{BRKLnnt}
\eeqa

\medskip

The search for processes with missing energy and branching ratios 
below $10^{-10}$ is definitely a very difficult challenge, but 
has been proved not to be impossible: two $K^+ \to \pi^+ \nu\bar\nu$ 
candidate events have been observed by the BNL-E787 experiment 
\cite{E787}. The branching ratio inferred from this result,
\beq
{\cal B} (K^+\to\pi^+ \nu\bar{\nu}) = \left( 1.57^{~+~1.75}_{~-~0.82} \right) \times 10^{-10}~,
\label{eq:E787}
\eeq
is consistent with SM expectations, although the error does not allow 
precision tests of the model yet. In a few years this result should
be substantially improved by the BNL-E949 experiment, whose goal is to  
collect about 10 events (at the SM rate).
In the longer term, a high-precision result on this mode will arise 
from the CKM experiment at Fermilab, which aims at a  measurement 
of ${\cal B}(K^+ \to\pi^+ \nu\bar{\nu})$ at the $10\%$ level \cite{Kettel_rev}.

Unfortunately the progress concerning the neutral mode is much slower.
No dedicated experiment has started yet (contrary to the $K^+$ case) 
and the best direct limit is more than four orders of magnitude above the 
SM expectation \cite{KTeV_nn}. An indirect model-independent upper bound  on
$\Gamma(K_L\to\pi^0\nu\bar{\nu})$ can be obtained by the isospin relation \cite{GN} 
\beq
\Gamma(K^+\to\pi^+\nu\bar{\nu})~=  \Gamma(K_L\to\pi^0\nu\bar{\nu}) +
\Gamma(K_S\to\pi^0\nu\bar{\nu}) 
\label{Tri}
\eeq
which is valid for any $s\to d \nu\bar\nu$ local operator of dimension 
$\leq 8$ (up to small isospin-breaking corrections).
Using the BNL-E787 result (\ref{eq:E787}), this implies 
${\cal B}(K_L\to\pi^0\nu\bar{\nu}) <  1.7 \times 10^{-9}~(90\%~{\rm CL})$.
Any experimental information below this figure can be translated into 
a non-trivial constraint on possible new-physics contributions to 
the $s\to d\nu\bar{\nu}$ amplitude. In a few years this goals should 
be reached by E931a at KEK: the first $K_L\to\pi^0\nu\bar{\nu}$ dedicated 
experiment. The only approved experiment that could 
reach the SM sensitivity on $K_L\to\pi^0\nu\bar{\nu}$ is KOPIO at BNL, whose goal 
is a SES of $10^{-13}$, or the observation of about 50 signal events (at the SM rate) 
with signal/background $\approx 2$ \cite{Kettel_rev}.

\section{Other Rare $K$ decays}
As far as theoretical cleanliness is concerned,
$K\to\pi\nu\bar{\nu}$ have essentially no competitors.
FCNC transitions where the neutrino pair is 
replaced by a charged-lepton pair have a
very similar short-distance structure. 
However, in these processes the size of long-distance 
contributions is usually much larger because of 
electromagnetic interactions. Only in few cases
(mainly in CP-violating observables) we can  extract the 
interesting short-distance information with reasonable accuracy 
(see Ref.~\cite{Kettel_rev,DI} and references therein).

\bigskip
\noindent
$\underline{K\to \pi \ell^+\ell^-}\ $
The  GIM mechanism of the $s \to d \gamma^*$ amplitude 
is only logarithmic \cite{GilmanW}.  
As a result, the  $K \to \pi \gamma^* \to \pi \ell^+\ell^-$ amplitude 
is completely dominated by long-distance dynamics 
and  provides a large contribution to the CP-allowed transitions  
$K^+ \to \pi^+ \ell^+ \ell^-$ and $K_S \to \pi^0 \ell^+ \ell^-$ 
\cite{EPR}. Rate and form factor of the charged mode 
have been measured with high accuracy  by  BNL-E865 \cite{E865}.
However, this information is not sufficient to predict the rate 
of the neutral mode. The latter is usually parameterized as 
${\cal B}(K_S \to \pi^0 e^+ e^-) = 5 \times 10^{-9} \times |a_S|^2$
where $a_S$ is a low-energy free parameter expected to be $\cO(1)$ \cite{DEIP}.  
The present experimental bound 
${\cal B}(K_S \to \pi^0 e^+ e^-) < 1.4 \times 10^{-7}$ \cite{NA48KS}
is still one order of magnitude 
above the most optimistic expectations, but a measurement or 
a very stringent bound on $|a_S|$ will soon arise from 
the $K_S$-dedicated run of NA48  and/or from the 
 KLOE experiment at Frascati.

Apart from its intrinsic interest, the determination of 
${\cal B}(K_S \to \pi^0 e^+ e^-)$ has important consequences on
the $K_L \to \pi^0 e^+ e^-$ mode. Here the long-distance part 
of the single-photon exchange amplitude is forbidden 
by CP invariance and the sensitivity to 
short-distance dynamics in enhanced. 
The direct-CP-violating part of the 
$K_L \to \pi^0 \ell^+ \ell^-$ amplitude is conceptually similar 
to the one of $K_L \to \pi^0 \nu \bar{\nu}$: it is calculable 
with high precision, being dominated by the top-quark contribution \cite{BLMM},
and is highly sensitive to non-standard dynamics.  
This amplitude interfere with the indirect-CP-violating 
contribution induced by  $K_L$--$K_S$ mixing, leading to \cite{DEIP}
\beq
{\cal B}(K_L \rightarrow \pi^0 e^+ e^-)_{\rm CPV}  =  10^{-12} \times 
\left[ 15.3 |a_S|^2 \pm  6.8 \frac{\displaystyle \Im \lambda_t}{\displaystyle 10^{-4}} 
|a_S| + 2.8 \left( \frac{\displaystyle \Im \lambda_t}{\displaystyle 10^{-4}}
\right)^2 \right] \label{eq:BKLee}
\eeq
where the $\pm$ depends on the relative sign between short- and 
long-distance contributions, and cannot be determined in 
a model-independent way. Given the present uncertainty on ${\cal B}(K_S \to
\pi^0 e^+ e^-)$, at the moment we can only set a rough upper limit
of $5.4 \times 10^{-10}$ on the sum of all the CP-violating 
contributions to this mode, to be compared with the direct 
limit of $5.6 \times 10^{-10}$ obtained by KTeV
at Fermilab \cite{KTeV_ee}.

An additional contribution to $K_L \to \pi^0 \ell^+ \ell^-$ 
decays is generated by the CP-conserving long-distance processes
$K_L \to \pi^0 \gamma^* \gamma^* \to \pi^0 \ell^+ \ell^-$ \cite{Sehgal}.
This amplitude does not interfere with the 
CP-violating one, and recent NA48 data on $K_L \to 
\pi^0\gamma\gamma$ (at small dilepton invariant mass)
indicate that it is very suppressed, 
with an impact on ${\cal B}(K_L \to \pi^0 e^+e^-)$ below the  
$10^{-12}$ level \cite{NA48gg}.

At the moment there exist no definite plans to improve the 
KTeV bound on ${\cal B}(K_L \rightarrow \pi^0 e^+ e^-)$. The future information 
on ${\cal B}(K_S \to \pi^0 e^+ e^-)$ will play a crucial role in this respect: 
if $a_S$ were in the range that maximizes the interference effect in 
(\ref{eq:BKLee}), we believe it would be worthwhile to start a 
dedicated program to reach sensitivities of $10^{-12}$. 

\bigskip
\noindent
$\underline{K_L \to \ell^+ \ell^-}\ $
Both $K_L \to \mu^+ \mu^-$ and  $K_L \to e^+ e^-$ decays are 
dominated by the two-photon long-distance amplitude
 $K_L \to \gamma^*\gamma^* \to \ell^+ \ell^-$.
The absorptive part of the  latter is determined to good accuracy by the two-photon 
discontinuity and is calculable with high precision 
in terms of the $K_L\to \gamma \gamma$ rate. On the other hand, 
the dispersive contribution of the two-photon amplitude 
is a source of considerable theoretical uncertainties. 

In the $K_L\to e^+ e^-$ mode the dispersive integral 
is dominated by a large infrared logarithm 
[$\sim \ln(m_K^2/m^2_e)$], the coupling of which 
can be determined in a model-independent way from  
$\Gamma(K_L\to \gamma \gamma)$. As a result, 
$\Gamma(K_L\to e^+ e^-)$ can be estimated with 
good accuracy but is almost insensitive to 
short-distance dynamics \cite{VP}.

The $K_L\to \mu^+\mu^-$ mode is certainly more interesting from 
the short-distance point of view. Here the two-photon long-distance 
amplitude is not enhanced by large logs and is almost comparable 
in size with the short-distance one, 
sensitive to $\Re \lambda_t$  \cite{BB}.
Actually short- and long-distance 
dispersive parts cancel each other to a good extent, since the total 
$K_L\to \mu^+\mu^-$ rate (measured with high precision 
by BNL-E871 \cite{E871}) is almost saturated by the absorptive 
two-photon contribution.

\begin{figure}[t] 
$$
\hskip -1.8 cm
\includegraphics[width=10.0cm,height=7.5cm]{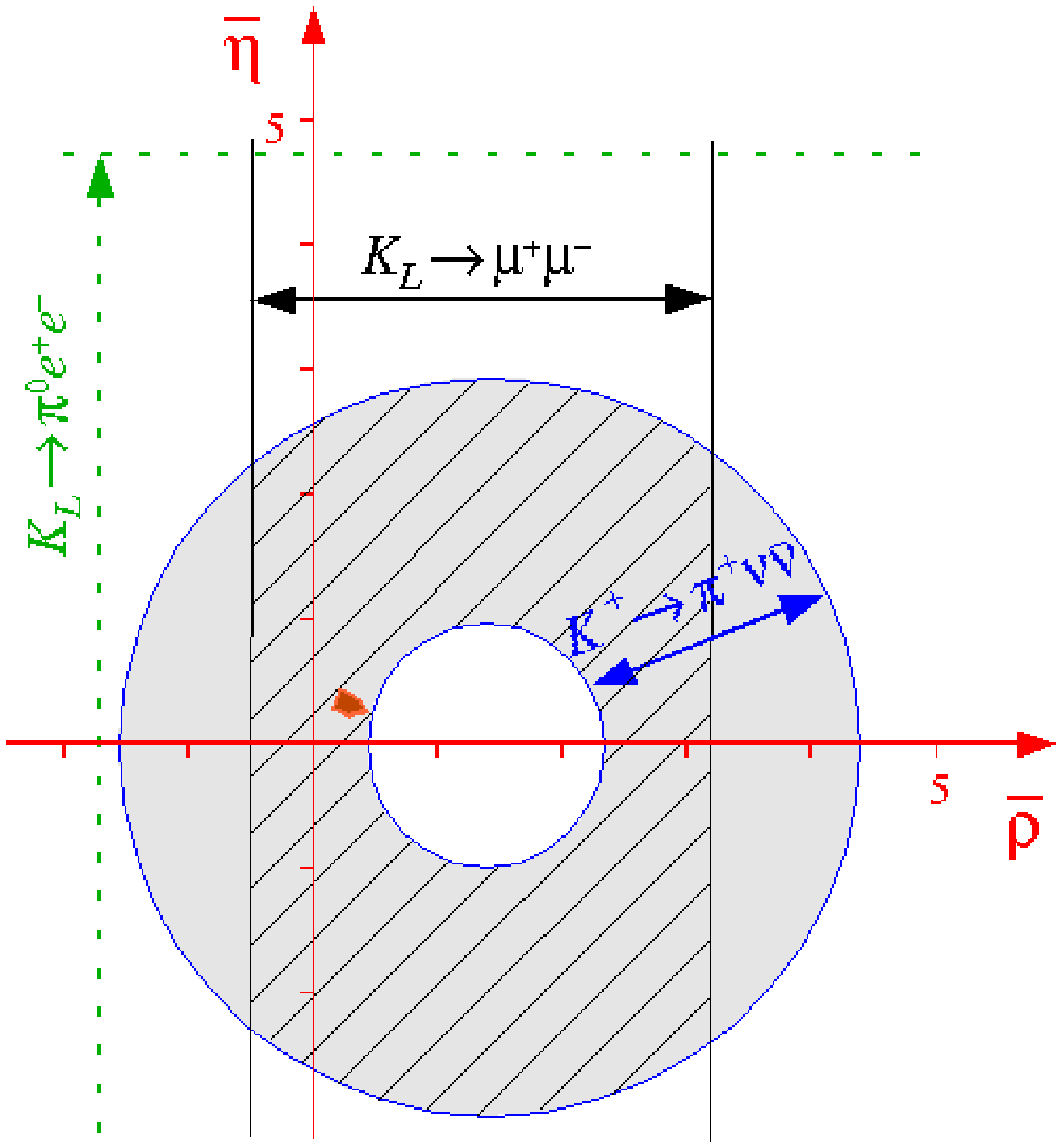}
\hskip -2.0 cm
\includegraphics[width=11.0cm,height=7.5cm]{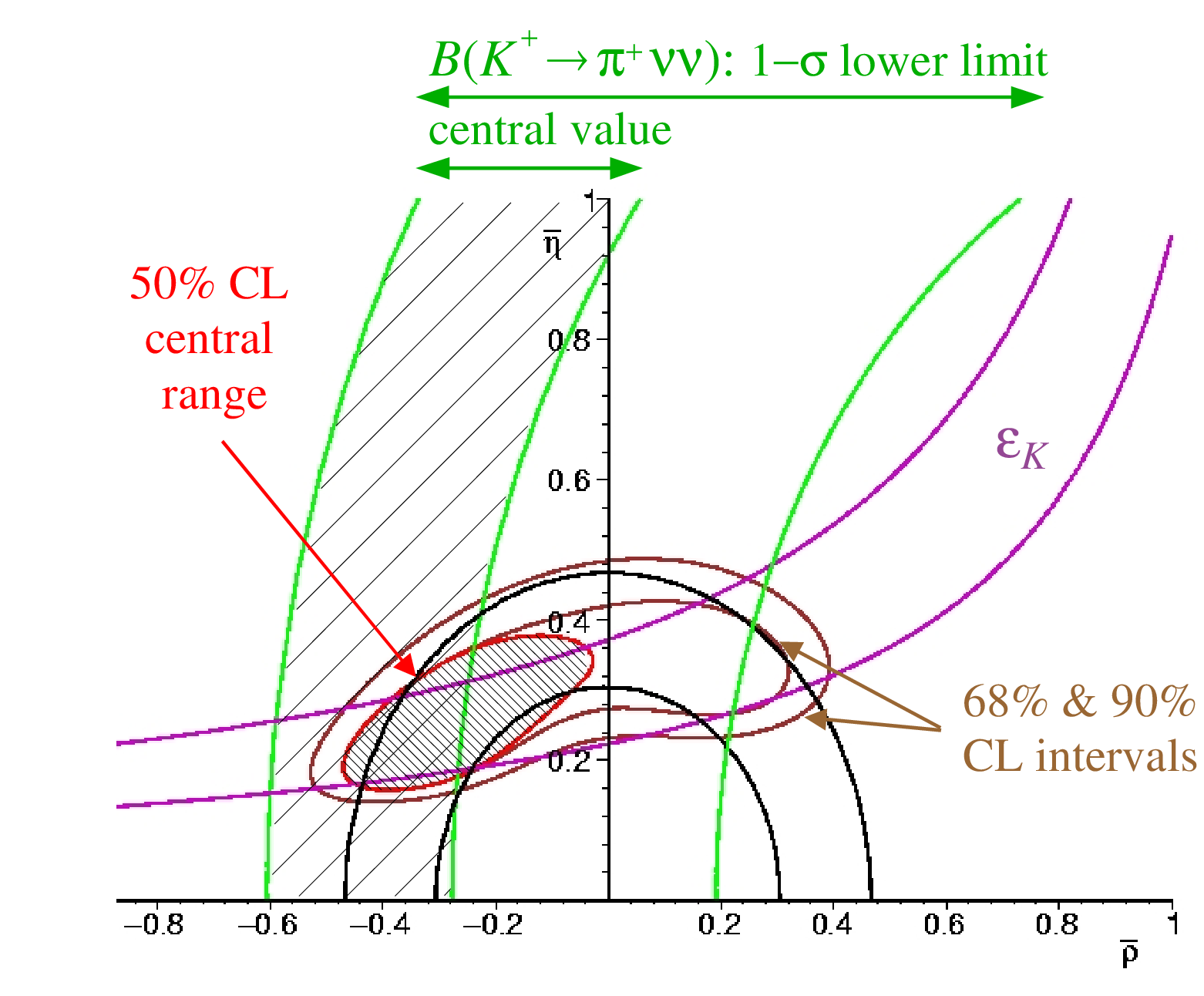}
\hskip -1.2 cm
$$
\vskip  0.5 cm
\caption{Left: present constraints in the $\bar\rho$--$\bar\eta$ plane from 
rare $K$ decays only (the small dark region close to the origin denotes 
the combined result of $B$-physics observables). Right: $\bar\rho$--$\bar\eta$ constraints
excluding observables sensitive to $B_d$-$\bar B_d$ mixing \cite{rising}. }
\label{fig:UT_rare}
\end{figure}

The accuracy on which we can bound the two-photon dispersive 
integral determines the accuracy of possible 
bounds on $\Re \lambda_t$. A partial control of the 
$K_L \to \gamma^* \gamma^*$ form factor, which rules the dispersive 
integral, can be obtained by means of $K_L
\to \gamma \ell^+\ell^-$ and $K_L \to e^+e^- \mu^+\mu^-$ spectra; 
additional constraints can also be obtained from model-dependent 
hadronic ansatze and/or perturbative QCD \cite{BMS,DIP}. 
Combining these inputs, significant 
upper bounds on $\Re \lambda_t$ (or lower bounds on $\bar \rho$) 
have recently been obtained \cite{E871,KTeV_alpha}. 
The reliability of these bounds has still to be fully investigated, 
but some progress can be expected in the near future. 
In particular, the extrapolation of the form factor 
in the high-energy region, which so far requires model-dependent 
assumptions, could possibly be limited by means of lattice 
calculations.

\section{Beyond the MFV hypothesis}
To conclude this  discussion about kaon physics, we summarize 
in Fig.~\ref{fig:UT_rare} (left) the present impact of rare $K$ decays in 
constraining the $\bar\rho$--$\bar\eta$ plane. As can be noted, 
the bounds from these modes are substantially  
less precise than those from $B$-physics. As a result, we are still 
far from precision tests of the main MFV prediction: the universality 
of non-standard effects in  $b \to d$,  $b \to s$ and $s \to d$ 
FCNC transitions \cite{noi}. On the other hand,
the comparison is already quite non-trivial concerning non-MFV scenarios. 
In particular, present rare-$K$-decay constrains put severe bounds 
on realistic scenarios with large new sources 
of flavour mixing in  $s \to d$ transitions
(see e.g. Ref.~\cite{BCIRS}).

Interestingly enough, non-MFV models with $\cO(1)$ effects in $s \to d$,
$b \to s$, and even $b \to d$ FCNC transitions are still far from
being excluded. As an example, in Fig.~\ref{fig:UT_rare} (right)
we show the result of a fit allowing arbitrary 
new-physics contributions to $B_d$--$\bar B_d$ mixing.
As can be noted, all remaining constraints are in good agreement; however,
the large central value of ${\cal B}(K^+\to \pi^+\nu\bar\nu)$ 
tends to flavour a CKM structure rather
different from the standard case \cite{rising}. 
This indication is not statistically 
significant yet, but it provides a good illustration 
of the main points of this discussion: there is still a lot to learn about
FCNC transitions and the measurements of $K \to \pi \nu\bar\nu$ rates 
provide a unique opportunity in this respect. 

\section*{Acknowledgments}
It is a pleasure to thank Giancarlo D'Ambrosio, Gian Giudice,
and Alesandro Strumia for useful discussions and an enjoyable 
collaboration. I am also grateful to the organizers of TH2002
for the invitation to this interesting and unique conference. 
This work is partially supported by IHP-RTN, 
EC contract No. HPRN-CT-2002-00311 (EURIDICE).

\frenchspacing
\footnotesize

\end{document}